\documentclass[12pt]{iopart}
\usepackage{braket}
\usepackage{amsopn}
\usepackage{amsfonts}
\usepackage{amssymb}
\usepackage{bbold}
\usepackage{graphicx}
\usepackage{color}
\usepackage[english]{babel}
\usepackage{placeins}
\usepackage{algorithm}
\usepackage{algorithmic}
\usepackage{subfigure}
\usepackage{enumerate}
\usepackage{psfrag,graphicx}
\usepackage{pict2e}
\usepackage{dcolumn}
\usepackage{bm}
\usepackage{latexsym}
\usepackage{epstopdf}

\usepackage{cancel}
\usepackage{appendix}
\definecolor{mygrey}{gray}{0.35}
\definecolor{myblue}{rgb}{0.2,0.2,0.8}
\definecolor{myzard}{cmyk}{0,0,0.05,0}
\definecolor{mywhite}{rgb}{1,1,1}
\definecolor{mywhite}{rgb}{1,1,1}
\definecolor{myred}{rgb}{1,0.,0.3}
\usepackage[colorlinks=true,citecolor=myblue,linkcolor=myred]{hyperref}
\def\cc{{\rm c}}

\def\rr{{\rm r}}

\def\ff{{\rm f}}
\def\adag{{a^{\dag}}}

\begin{document}
\title{Dissipative Josephson effect in coupled nanolasers}

\author{Samuel Fern\'andez-Lorenzo}
\address{OpenSistemas, Calle Salvatierra 4, 28034 Madrid, Spain}
\address{Department of Physics and Astronomy, University of Sussex, Falmer, Brighton BN1 9QH, UK}

\author{Diego Porras}

\address{Instituto de F\'{\i}sica Fundamental IFF-CSIC, Calle Serrano 113b, Madrid 28006, Spain}
\address{Department of Physics and Astronomy, University of Sussex, Falmer, Brighton BN1 9QH, UK}
\ead{diego.porras@csic.es}

\begin{abstract}
Josephson effects are commonly studied in quantum systems in which dissipation or noise can be neglected or do not play  a crucial role. 
In contrast, here we discuss a setup where dissipative interactions do amplify a photonic Josephson current, opening a doorway to dissipation-enhanced sensitivity of quantum-optical interferometry devices. 
In particular, we study two coupled nanolasers subjected to phase coherent drivings and coupled by a coherent photon tunneling process. 
We describe this system by means of a Fokker-Planck equation and show that it exhibits an interesting non-equilibrium phase diagram as a function of the coherent coupling between nanolasers. 
As we increase that coupling, we find a non-equilibrium phase transition between a phase-locked and a non-phase-locked steady-state, in which phase coherence is destroyed by the photon tunneling process. 
In the coherent, phase-locked regime, an imbalanced photon number population appears if there is a phase difference between the nanolasers, which appears in the steady-state as a result of the competition between competing local dissipative dynamics and the Josephson photo-current.
The latter is amplified for large incoherent pumping rates and it is also enchanced close to the lasing phase transition.
We show that the Josephson photocurrent can be used to measure optical phase differences. In the quantum limit, the accuracy of the two nanolaser interferometer grows with the square of the photon number and, thus, it can be enhanced by increasing the rate of incoherent pumping of photons into the nanolasers.
\end{abstract}

\maketitle
\setcounter{tocdepth}{2}
%\begingroup
%\hypersetup{linkcolor=black}
%%\tableofcontents
%\endgroup

\section{Introduction}

%%%%%%%%%%%%%%%%%%%%%%%%%%%%%%%%%%%%%%%%%

%
Among the striking effects of quantum coherence, the Josephson effect is one of widest used in nowadays technologies \cite{barone82}. 
To name a few examples, the SQUID (superconducting interference device) is one of the oldest and most sensitive magnetic sensors \cite{jaklevic65,fagaly06}, and Josephson junctions are integral building blocks to construct artificial two-level systems in quantum information processing \cite{nakamura99,martinis02}.
In precision metrology, the Josephson effect has been used as a practical standard of voltage and the elementary charge, $e$ \cite{taylor67,field73}. 

The Josephson effect occurs when two quantum systems having both well-defined quantum phases, $\phi_{1,2}$, are weakly coupled so that quantum tunneling is enabled between them. It is manifested as a net current, 
$$I=I_{\rm c}\sin(\Delta\phi),$$
between the two subsystems depending on the phase difference, $\Delta\phi = \phi_1 - \phi_2$. 
This effect was discovered by Josephson, who predicted a macroscopic electric current along a superconducting tunnel junction \cite{josephson62,anderson63}. 
Since then, extensions of these ideas have been proposed and tested in several platforms, such as 
Bose-Einstein Condensates \cite{Zapata97pra,smerzi97,williams99,giovanazzi00,albiez05,levy07}, 
superfluid $^{3}{\rm He}$ \cite{pereverzev97}, 
photonic \cite{gerace09,ji09,khomeriki06} and 
optomechanical systems \cite{larson11} 
and polaritons \cite{abbarchi13,shelykh08}. 
While the effect of dissipation has been studied  in some cases \cite{gerace09}, it has generally been considered as detrimental for the observation of the Josephson current and its applications.
%%%%%%%%%%%%%%%%%%%%%%%%%%%%%%%

\begin{figure}[h!]
  \includegraphics[width=\textwidth]{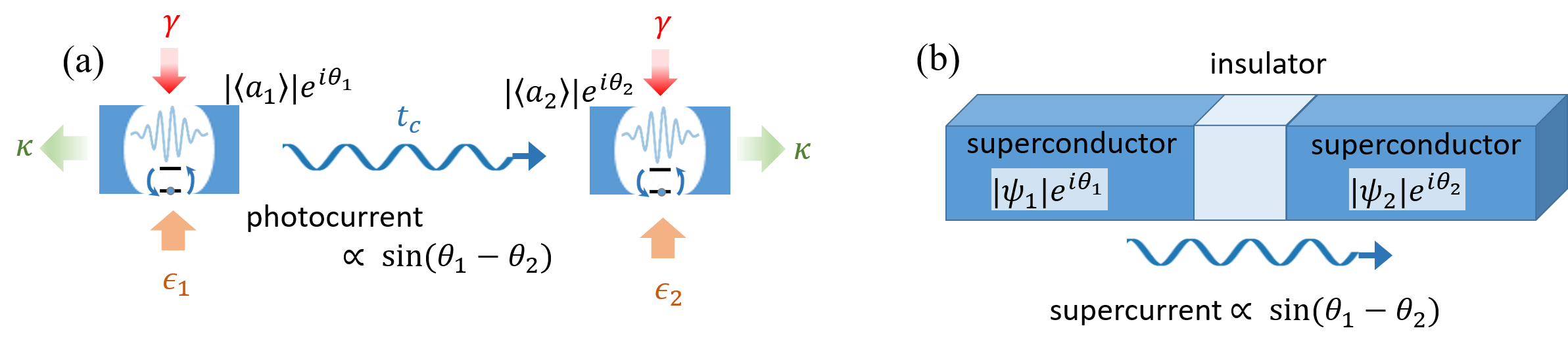}
  \caption{(a) General scheme: two single-qubit lasers are coupled by a coherent photon hopping term with rate $t_{\rm c}$. Each nanolaser is and subjected to incoherent qubit pumping with rate $\gamma$, photon loss with rate $\kappa$, and a periodic driving field with amplitude $\epsilon_{1,2}=|\epsilon|e^{i\phi_{1,2}}$. A Josephson photocurrent is generated if there is a finite difference between the optical phases, $\theta_1$, $\theta_2$, at each nanolaser. (b) Conventional Josephson effect: an electric current is generated across a junction separating two superconductors with different phases.}
  \label{fig:fig1} 
\end{figure}

In this work we unveil a dissipative Josephson in a fully photonic setup consisting of two coupled nanolasers. We show that this system presents interesting non-equilibrium phases and it also has an exciting outlook for applications in quantum metrology and sensing. 
In particular, we introduce a model of an interferometer consisting of two single qubit lasers coherently coupled through a photon tunneling term, as depicted schematically in Fig. \ref{fig:fig1}. 
Nanolasers are probably the most fundamental example of active dissipative systems with a non-trivial non-equilibrium phase diagram.
They can be implemented in several platforms such as photonic 
\cite{mckeever03, astafiev07nat, nomura10natphys,liu15sci}, 
plasmonic \cite{lu12sci}, or 
nano-mechanical systems \cite{vahala09}. 
A natural extension from the single nanolaser model into the many-body regime arises when we consider networks of local nanolasers coupled by means of photon tunneling terms of the form 
$H_{t} = t (a^\dagger_1 a_2 + a_1 a^\dagger_2),$ \cite{Hartmann16,houck12natphys}.
Here, interesting phenomena may appears as a result of the interplay between coherent tunneling and on-site non-linear dissipative terms. In our setup those nonlinear dissipative terms are responsible for sustaining the quantum coherence that generates the Josephson effect and its possible applications in interferometry.

This article presents the following results:
(i) We derive a semi-classical description of a dissipative interferometer consisting of two nanolasers coupled through a coherent photon tunneling process and subjected to coherent drivings with phase difference, $\Delta \phi$. 
(ii) We identify two limiting regimes of the steady-state of this system. If the coherent coupling is small, the lasers are phase locked to each individual coherent driving. As we increase the coherent coupling between nanolasers, the system goes through a non-equilibrium phase transition into a steady-state in which phase coherence is lost.
(iii) We present analytical and numerical evidence of the existence of a photonic Josephson current between the two nanolasers.  
This current causes a photon number imbalance in the photonic steady-state proportional to $\sin (\Delta\phi)$. 
(iv) We analyse the performance of this system as an interferometer and we show that there is an amplification effect by which the accuracy grows as $\sqrt{n}$ in the shot noise limit, with $n$ the number of photons in the nanolasers. 
Furthermore, the accuracy is optimal close to the critical point of the lasing phase transition.

%%%%%%%%%%%%%%%%%%%%%%%%%%%%%%%%%%%%%%%%%%
\section{Theoretical description of coherently coupled nanolasers}
\subsection{Dimer of single-qubit lasers}
Our system consists of two single-qubit lasers coupled by coherent photon tunneling. 
This scheme can be implemented in several setups including circuit QED (for example 
using the ideas proposed in \cite{navarrete14prl}) and trapped ion phonon lasers \cite{vahala09}. 
The discussion below is focused on the case of optical nanolasers, however, our results are independent of any particular realization in photonics, vibronics or optomechanics.

Each nanolaser laser consists of a two-level system (qubit) with levels $|g\rangle$ and $|e\rangle$, coupled on resonance with a cavity mode through a Jaynes-Cummings type interaction. 
The two-level system is incoherently pumped with a rate $\gamma$ and the photonic mode has a decay rate $\kappa$. 
In addition, the cavity interacts with a weak coherent driving on resonance with the photonic mode. 
The two nanolasers are connected by a photon coherent tunneling term that couples the photonic cavities.
In an interaction picture rotating at the mode frequency, the following master equation for the system density matrix, $\rho$, captures the complete quantum dynamics of the system (we use units such that $\hbar = 1$),
\begin{equation} 
\dot{\rho} = -i \ [H,\rho]+
\sum_{j=1,2} \left(\mathcal{L}_{\{\sigma_j^+ , \gamma\}} \left( \rho \right) + \mathcal{L}_{\{a_j , \kappa\}}\left( \rho \right) \right) .
\label{Liouvillian}
\end{equation}
We use the notation 
$\mathcal{L}_{\{O , \Gamma\}}(\rho)= \Gamma (2 O\rho O^{\dag}-O^{\dag}O\rho - \rho O^{\dag}O)$ 
for Lindblad super-operators. The Hamiltonian in Eq. (\ref{Liouvillian}) is
\begin{eqnarray}
H = g \sum_{j = 1,2} \left( \sigma_j^+ a_j+ a^\dagger_j\sigma_j^- \right) 
+ \sum_{j = 1,2} \left(  \epsilon_j^* a_j  + \epsilon_j a_j^{\dagger} \right) 
+ H_{\rm c}.
 \label{Hamiltonian} \nonumber
\end{eqnarray}
The first term represents the qubit-field coupling, with strength $g$. 
The second one describes external driving terms acting on the nanolasers, 
with driving amplitudes $\epsilon_j = |\epsilon|e^{i\phi_j}$. 
In this work we will consider that both driving amplitudes have the same strength, $|\epsilon|$, but there may be a phase difference, 
$\Delta \phi = \phi_2 - \phi_1$. Finally, the last term represents the coherent photon tunneling term, with amplitude $t_{\rm c}$,
\begin{eqnarray}
H_{\rm c}  = - t_{\rm c} \left( a^{\dag}_1 a_2+a^{\dag}_2 a_1 \right).
 \label{Ht} 
\end{eqnarray}
This coherent coupling occurs in systems of single-mode nano-cavity arrays 
\cite{Hartmann08,Hartmann16}, superconducting circuits \cite{houck12natphys} or nano-mechanical systems such as trapped ions (where it takes the form of a phonon-tunneling term, see \cite{Porras04prl2,Haze12pra}). 
The minus sign is added to the coupling so that for positive $t_{\rm c}$ the lowest energy mode is the symmetric or center of mass mode. 
At first sight, we would expect a term like 
Eq. (\ref{Ht}) to induce some synchronization of the phase between nanolasers, however, we will prove later on that this intuition is wrong and, actually, strong tunneling constants, $t_\cc$ tend to destroy the quantum coherence in the system.

The steady-state of each nanolaser is governed by the parameter
\begin{equation}
C_{\rm p} = \frac{g^2}{\kappa \gamma},
\end{equation}
such that, for $C_{\rm p} < 1$, nanolasers are in a non-lasing steady-state, whereas $C_{\rm p} > 1$ is the lasing phase, with a number of photons that roughly scales like $\gamma/\kappa$, as we show below.
In this paper, we will assume that the local nanolaser dynamics is much faster than the photon tunneling term, and in particular, $t_{\rm c} \ll \gamma$, such that we can safely assume that the lasing transition stays at $C_{\rm p} = 1$.
The mean-field prediction for the number of photons is \cite{BreuerPet}, 
\begin{eqnarray}
n_{\rm mf} &=& 0,   \hspace{1cm} C_{\rm p} < 1
\nonumber \\
n_{\rm mf} &=& \frac{1}{2} \frac{C_{\rm p - 1}}{C_{\rm p}} \frac{\gamma}{\kappa},   
\hspace{1cm} C_{\rm p} \geq 1
\label{n0}
\end{eqnarray}
From Eq. (\ref{n0}), we learn that the parameter that determines the effective size of our 
nanolaser model is actually the ratio $\gamma/\kappa$, 
which determines any finite-size scaling effects and plays a role that is analogous to the number of sites in a quantum lattice model 
(see \ref{App:AppendixB}).

\subsection{Effective non-linear photonic master equation}

We expect the most interesting physics occurring in a lasing regime of large photon 
numbers in which each nanolaser can be approximated by a self-sustained 
quantum oscillator. 
This regime may be attained for a strong pumping of the qubit such that 
\begin{equation}
\gamma \gg g, \kappa, |\epsilon|, t_{\rm c}  .
\end{equation}
All along this paper we will be working in this regime, in which we can simplify our model by adiabatically eliminating
the qubit dynamics \cite{Mandel95}. This step will ultimately allow us to obtain a semiclassical description in terms of a Fokker-Planck equation. 
After the qubit's adiabatic elimination we get the following master equation 
(see \ref{App:AppendixB} for details), 
\begin{eqnarray} 
	& & \dot{\rho}_{\rm f} 
	= i \ t_{\rm c}
	 [a^{\dag}_1 a_2+a^{\dag}_2 a_1,\rho_\ff]
	 - i \sum_{j=1,2}[\epsilon_j^* a_j  + \epsilon_j a_j^{\dag},\rho_\ff]   
+ \sum_{j = 1,2} \mathcal{L}_{\{a_j, C\}}(\rho_\ff) 	 
\label{adiabatic} 
\\
&+& \sum_{j=1,2}\left(\mathcal{L}_{\{a^{\dag}_j, A\}}(\rho_\ff) 
+ 
\frac{1}{2} \mathcal{L}_{\{a_j a^{\dag}_j, B\}}(\rho_\ff) - 
\frac{1}{2} \mathcal{L}_{\{(a^{\dag}_j)^2, B\}}(\rho_\ff) +
\frac{1}{2} \mathcal{L}_{\{(a^{\dag}_j), B \}}(\rho_\ff)
\right) ,
\nonumber
\end{eqnarray}
with the coefficients:
\begin{equation}
A = \frac{g^2}{\gamma}, \ \ B = 2 \frac{g^4}{\gamma^3}, \ \ C = \kappa.  
\end{equation}
$\rho_{\rm f} = {\rm Tr}_{\rm qubit} \left( \rho \right)$ is the reduced density matrix of the photonic 
subsystem, obtained after tracing out the qubit. 
The last two terms in Eq. (\ref{adiabatic}), proportional to the coefficient $B$, account for the non-linear matter-light interaction and they are ultimately responsible for the non-equilibrium phase transition into the lasing phase.

Eq. (\ref{adiabatic}) is strictly valid only below the critical point, 
$C_{\rm p} < 1$, and 
slightly above it, $C_{\rm p}\gtrsim 1$, since it has been obtained under the assumption of total qubit population inversion (see \ref{App:AppendixB}). 
To quantify better the regime of validity of this approximation we can calculate the ground-state population of the qubit in the steady-state, 
$\langle \sigma^+_j \sigma^-_j \rangle_{\rm ss}$,
and check whether it can really be neglected. 
Actually, by using the equations for the single-qubit case, we find
\cite{Lorenzo17},
\begin{eqnarray}
\langle \sigma^+_j \sigma^-_j \rangle_{\rm ss}  =
{\rm Tr} \left( \rho_{\rm ss} \right)
= \left( \frac{g}{\gamma} \right)^2 
\left( 1 + \langle a^\dagger_j a_j \rangle_{\rm ss} \right)
&\approx& \frac{1}{2} \frac{C_{\rm p} - 1}{C_{\rm p}}, \ \ \ C_{\rm p} > 1
\nonumber \\
&\approx& 0  ,  \ \ \ \ \ \ \ \ \ \ \ \ \ \ C_{\rm p} < 1 .
\label{q.population}
\end{eqnarray}
We confirm that condition 
$\langle \sigma^+_j \sigma^-_j \rangle_{\rm ss} \ll 1$ is met close to the phase transition at $C_{\rm p} \gtrsim 1$, and in the non-lasing phase ($C_{\rm p} < 1$).
It is, however, 
highly desirable to extend Eq. (\ref{adiabatic}) well into the lasing phase 
($C_{\rm p} > 1$), 
to fully understand the systems's behavior.
This can be done with a proper renormalization of the coefficient that takes into account 
the neglected terms in the adiabatic elimination. 
As we show in \ref{App:AppendixB}, this procedure amounts to replacing 
\begin{equation}
B\rightarrow B/C_{\rm p}\equiv B_{\rm r},
\end{equation}
in Eq. (\ref{adiabatic}). The new parameter $B_{\rm r}$ includes the effect of processes of higher order in $g$, and it ensures the right prediction on the photon number in the steady-state.

\subsection{Semiclassical Fokker-Planck equation}
Although we are dealing with only two nanolasers, the solution of Eq. (\ref{adiabatic}) is computationally demanding because a very high number of photonic Fock states must be included in any exact numerical calculation. 
However, in the limit of high-photon numbers an analytical approach based on phase-space methods can be used to further simplify Eq. (\ref{adiabatic}). 
Concretely, we shall employ the \textit{Glauber-Sudarshan P} representation \cite{Mandel95} of the effective master equation. This representation is defined as the pseudo-probability distribution satisfying 
\begin{equation}
\label{Glauber-Sudarshan}
\rho_{\rm f}(t) = 
\int d^2\alpha 
P(\alpha,\alpha^*,t) |\alpha\rangle\langle \alpha |,
\end{equation} 
where $|\alpha\rangle$ is the coherent state $|\alpha\rangle=\exp{(\alpha 
a^{\dag}-\alpha^* a)}|0\rangle$. 
The function $P(\alpha,\alpha^*)$ plays the role of a classical probability distribution over $|\alpha\rangle \langle\alpha|$, with the normalization condition
$\int d^2 \alpha P(\alpha,\alpha^*) = 1$. Expectation values of normal ordered operators can be evaluated with the identity,
\begin{equation}
    \langle (a^\dagger)^p a^q \rangle = \int d^2 \alpha \ (\alpha^*)^p \alpha^q 
    P(\alpha,\alpha^*).
\end{equation}

The substitution of Eq. (\ref{Glauber-Sudarshan}) into Eq. (\ref{adiabatic}) transforms the master equation into a 
Fokker-Planck equation for $P(\alpha,\alpha^*,t)$ \cite{Risken84}, see  \ref{App:AppendixD}. We can achieve further simplification by working with polar coordinates,
\begin{equation}
\alpha_j = r_j e^{i\theta_j} \ \ (j=1,2) .
\end{equation}
The radial components, $r_j$, are related to the photon number observable in each cavity, 
\begin{equation}
\langle r_{j}^2 \rangle   
= \int d^2 \alpha |\alpha_j|^2 P(\alpha,\alpha^*,t) 
=  \langle n_j \rangle \  (\equiv \langle a^\dagger_j a_j \rangle) . 
\end{equation}
 In the last equation and along the rest of this work, we will understand $\langle O \rangle$ as refering to both quantum average, or average with respect to the semiclassical distribution, $P$, depending on whether $O$ is an operator or a Fokker-Planck variable.
 The key to simplifying the Fokker-Planck equation is the observation that the fluctuations in the radial components, $r_j$,  may be neglected as long as we are deep enough in the lasing regime ($C_{\rm p} > 1$ and 
 $\langle n_j \rangle \gg 1$). Thereby we can assume that radial variables remain close to their steady-state values $r_j\approx r^0_j$. In this case we can trace out the radial variables and consider a reduced description in terms of phase variables, $\theta_{j}$. Formally, this is accomplished by assuming a factorized 
 $P(\alpha_1,\alpha_2) \approx R(r_1) R(r_2) P_\theta(\theta_1,\theta_2)$.
Each $R(r_j)$ is a Gaussian distribution properly normalized around $r^0_j$ (average value of $r_j$), which corresponds to the radial probability distribution of each nanolaser in the lasing regime.
This procedure is discussed in details in \ref{App:AppendixD}, and leads to and equation that depends on the phases of the nanolasers only,
\begin{eqnarray} 
\frac{\partial P_\theta}{\partial t} 
&=& \frac{A}{2}
\sum_{j=1,2} \frac{1}{n^0_j} \frac{\partial^2 P_\theta}{\partial \theta_j^2} 
\nonumber \\
&+& \sum_{j=1,2} 
\frac{\partial}{\partial \theta_j} \left( -t_{\rm c} \frac{r^0_{j+1}}{r^0_j} \cos(\theta_{j+1}-\theta_j) + \frac{|\epsilon|}{r^0_j}\cos(\theta_j-\phi_j)\right) P_\theta,
\label{PFokker2.0}
\end{eqnarray}
in which $n^0_j=(r^0_j)^2$ stands for the steady-state average number of bosons at site $j$. 
In the lasing regime and in the absence of tunneling, this quantity is independent of the site and is given by $n_j^0 = n_0$, with
\begin{equation}
n_0 \equiv \frac{A-C}{B_{\rm r}} = \frac{1}{2} \frac{C_{\rm p} - 1}{C_{\rm p}} 
\frac{\gamma}{\kappa} .
\label{n0.2}
\end{equation}

In the homogeneous case we have $r^0_j = r_0$, which leads to a homogeneous Fokker-Planck equation for the phases,
\begin{eqnarray} 
&& \frac{\partial P_\theta}{\partial t} 
= \nonumber \\  
&& \sum_{j = 1,2} 
\left( D_{\rm p}  \frac{\partial^2 P_\theta}{\partial \theta_j^2}  +
 \frac{\partial}{\partial \theta_j} \left(   - t_{\rm c} \cos(\theta_{j+1} - \theta_j) + \frac{|\epsilon|}{r_0}\cos(\theta_j-\phi_j)\right) \right) P_\theta ,
\label{PFokker2}
\end{eqnarray}
representing a uniform photon density together with a single phase diffusion rate, 
\begin{equation} 
D_{\rm p} \equiv \frac{A}{2 n_0}.
\end{equation}
We will see later that this picture has to be corrected to account for photon imbalance induced by the  Josephson current between nanolasers.

Eq. (\ref{PFokker2}) is one of the most important results for our work. We remark that the novel element in this equation is the coupling between phases induced by $t_{\rm c}$. 
Actually, this equation is closely related to the dissipative Kuramoto model, however, in our phase model there is a coherent coupling term which differs from the usual dissipative couplings considered in  coupled laser or synchronization models.
This will have severe consequences in the non-equilibrium phase diagram of the model, as we see below.

%%%%%%%%%%%%%%%%%%%%%%%%%%%%%%%%%%%%%%%%%%%%%%%%%%%
\section{Dissipative phase transition induced by coherent photon tunneling}

We investigate now the effect of the coherent photon coupling in the steady-state of our Fokker-Planck equation (\ref{PFokker2}) and show that, surprisingly, it does not lead to any synchronization effect between the nanolasers. On the contrary a coherent photon tunneling process leads to the loss of quantum coherence in the system. 

\subsection{Phase-locked and non-phase-locked steady-states}
To simplify the discussion we consider first an homogeneous driving with $\phi_1 = \phi_2 = \phi$, and $r^0_1 = r^0_2 = r_0$. 
In Eq. (\ref{PFokker2}) there are two limiting cases: 

\begin{figure}[h]
	\includegraphics[width = 0.8\textwidth]{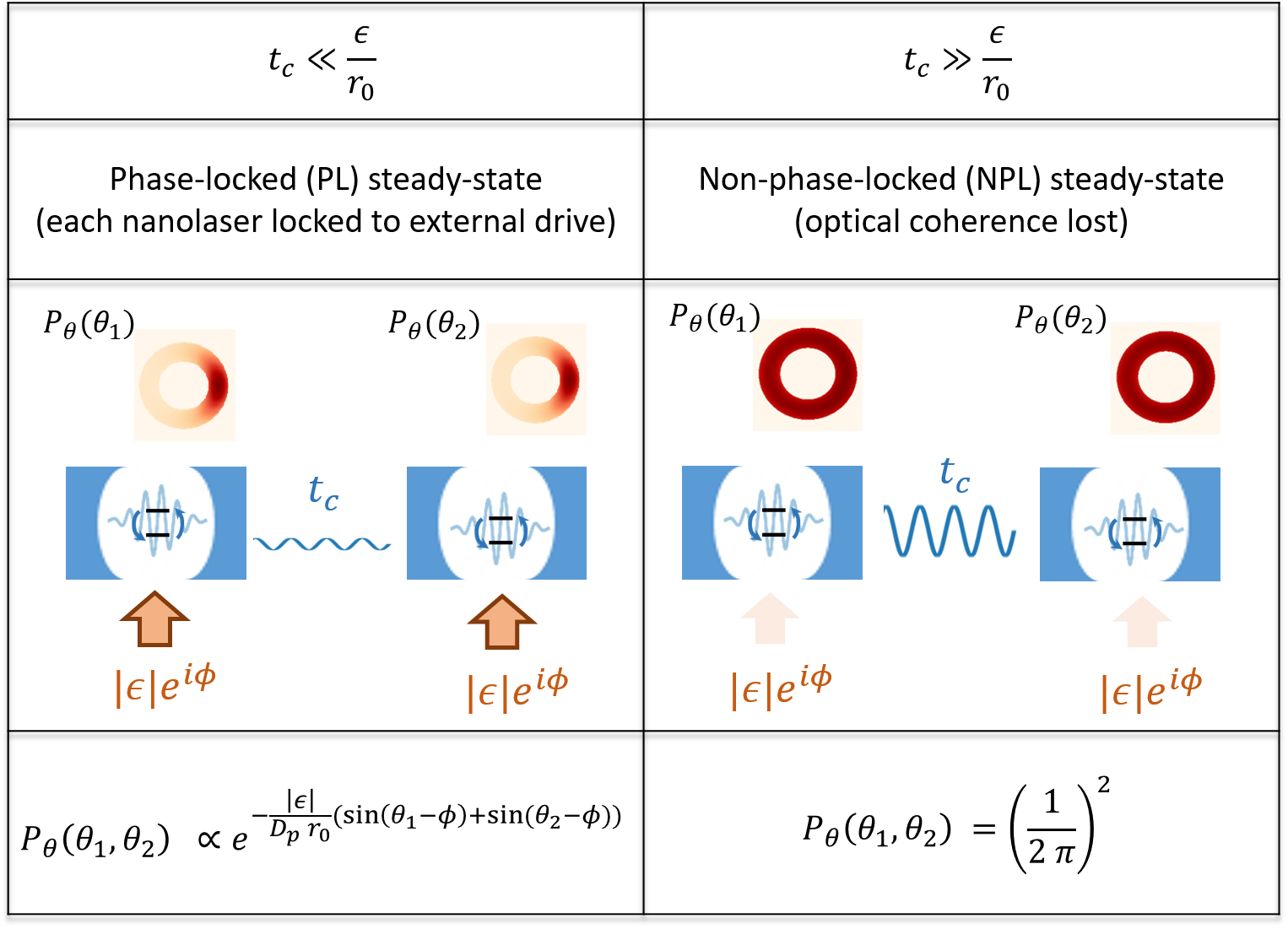}
	\centering
	\caption{Non-equilibrium phases of the nanolaser dimer in the lasing regime. 
	In the PL phase (left), nanolasers are phase locked to the external coherent fields. The semiclassical probability distribution is strongly concentrated around  $\theta_j = \phi_j - \pi/2$. In the NPL phase the photon tunneling between nanolasers is the dominant process, and it induces a total loss of phase coherence in the system.}
	\label{desynchronization_scheme} 
\end{figure}

\begin{enumerate}[(i)]
\item  $\epsilon/r^0 \gg t_{\rm c}.$ In this limit we expect that the system is well approximated by two independent single qubit lasers phase-locked to the driving fields, as explored in Ref. \cite{Lorenzo17}. 
In this case the Fokker-Planck equation can be exactly solved and we get
\begin{equation}
P_\theta \left(\theta_1, \theta_2 \right) 
\propto e^{- \frac{|\epsilon|/r_0}{D_{\rm p}} \sum_j \sin(\theta_j - \phi)} .
\end{equation}
Hence, here we find non-zero coherences 
$|\langle a_j\rangle|\neq 0$. We will refer to this steady-state as a phase-locked (PL) phase.

\item $t_{\rm c} \gg |\epsilon|/r^0$, where we expect that the coherent tunneling dominates the system's dynamics. 
To understand this limit it is useful to study 
Eq. (\ref{PFokker2}) with $|\epsilon| = 0$,
\begin{equation} \label{PFokker3}
\frac{\partial P_\theta}{\partial t} 
= D_{\rm p}
\left( 
\frac{\partial^2 P_\theta}{\partial \theta_1^2}  
+ 
\frac{\partial^2 P_\theta}{\partial \theta_2^2}  
\right)
- t_{\rm c} \left( \frac{\partial}{\partial\theta_1}  +  \frac{\partial}{\partial\theta_2} \right)
 \cos(\theta_2 - \theta_1) P_\theta .
\end{equation}
%
%where we have imposed the homogeneity condition and defined the phase diffusion rate,
%\begin{equation} 
%D_{\rm p} = A/(2 n_0).
%\end{equation}
Eq. (\ref{PFokker3}) can be solved by a change of variables to collective coordinates, 
$\theta_+ = (\theta_1 + \theta_2)/\sqrt{2}$, 
$\theta_- = (\theta_2 - \theta_1)/\sqrt{2}$.
In the new variables we find the steady-state solution, 
\begin{equation}
P_\theta = 1/(2 \pi),
\end{equation}
with vanishing coherences, $\langle a_j \rangle = 0$. 
Furthermore, in this limiting case, the two-point correlation $G = \langle a^{\dag}_1a_2\rangle$ also becomes zero, which implies the absence of any synchronization between the modes \cite{amitai17}. 
We will refer to a steady-state where $t_{\rm c}$ is dominant, and the system does not retain any coherence, as non-phase-locked (NPL) phase.
\end{enumerate}
We arrive to the somehow counter-intuitive conclusion that the presence of a coherent photon tunneling term does not induce any correlations between nanolasers. This situation differs strongly from the case of a dissipative coupling as induced, e.g. by incoherent tunneling through evanescent modes or an intermediate lossy cavity \cite{Fernandez18}, in which coupling does induce a phase correlation between nanolasers.

\subsection{Phase transition between PL and NPL steady-states: driving with homogeneous phases ($\phi_1 = \phi_2$)}

What happens in the intermediate regime between the PL and NPL steady-states identified above? 
To address this question, we assume that we have a constant driving term, $\epsilon$, 
and we increase the tunneling from $t_{\rm c} \ll \epsilon/r^0$ to 
$t_{\rm c} \gg \epsilon/r^0$. 
In particular, we are interested to know whether a dissipative phase transition separates the two phases. 
The coherences $|\langle a_j \rangle|$ can be used as the order parameter to distinguish between the PL and NPL steady-states.
We also need to define a valid thermodynamic limit to establish the existence of critical properties.
Even though this is a two-site system, a thermodynamic limit is obtained by letting the number of photons in the steady state, 
$n_0$, play the role of the system size \cite{hwang15,Lorenzo17}. 
The number of photons is essentially regulated by the ratio $\gamma/\kappa$ as shown in Eq. (\ref{n0}), so that the thermodynamic limit will be reached in a limit of strong pumping, $\gamma \gg \kappa$.
We have solved numerically Eq. (\ref{PFokker2}) by discretizing the angular variables, $\theta_1$, $\theta_2$, in a number of values, $n_{\rm c}$, running from $0$ to $2 \pi$. 
This procedure allows us to express $P_{\theta}$ as a vector and $\partial P_\theta / \partial t$ as a non-Hermitian matrix, and to calculate numerically the steady-state solution,
$\partial P_\theta / \partial t = 0$.
We expect that this numerical method is accurate as long as 
\begin{equation}
\frac{n_{\rm c}}{2 \pi} \gg \frac{1}{P_\theta} \frac{\partial P_\theta}{\partial \theta_j},
\end{equation}
or, equivalently, that the angular probability distribution function does not change much within a phase interval $\Delta \theta = 2 \pi /n_{\rm c}$. We have checked that in all the calculations shown in this work the numerical results have converged for the values of $n_{\rm c}$ used. Numerical calculations allow us to find the reduced probability distributions, 
\begin{equation}
P_1(\alpha_1) = \int d\alpha_2 P(\alpha_1,\alpha_2) \approx 
R(r_1) \int d\theta_2 P_\theta (\theta_1,\theta_2), 
\end{equation}
where in the last terms we have re-expressed the reduced probability for $\alpha_1$ in polar coordinates.
An analogous definition holds for $P_2(\alpha_2)$.
\begin{figure}[h!]
	\includegraphics[width=\textwidth]{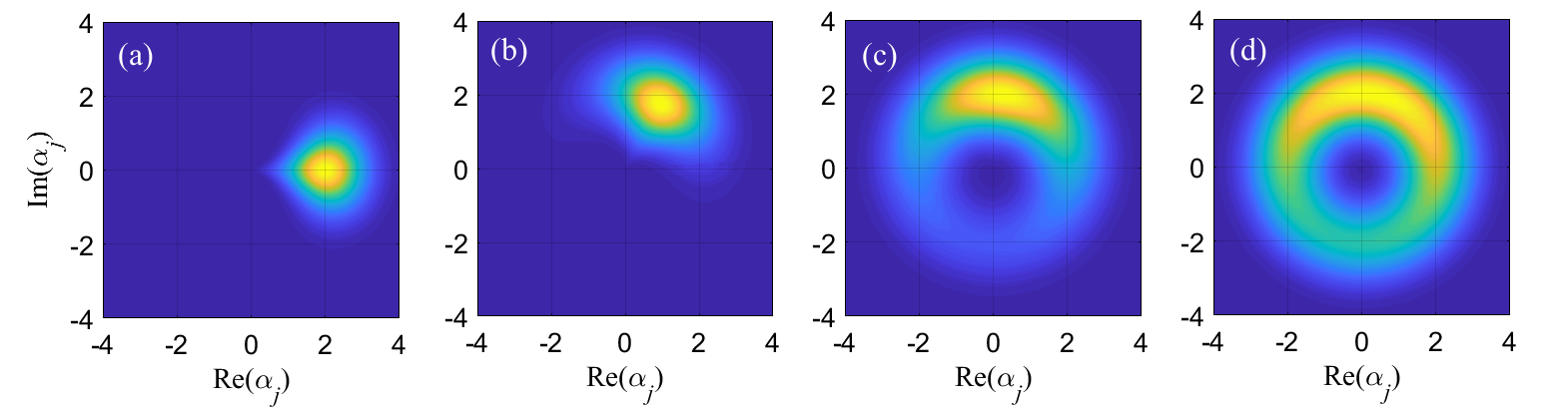}
	\caption{Reduced probability distribution $P_j(\alpha_j)$ (with $j = 1$ or $2$), calculated with the Fokker-Planck equation, Eq. (\ref{PFokker2}) with:  
	(a) $t_\cc = 0$, (b) $t_\cc = 1$, (c) $t_\cc = 2$, (d) $t_\cc = 5$. All calculations are carried out with $n_\cc = 100$, $D_{\rm p} = 0.1$, $\epsilon = 1$, and $\phi_{1,2}  = \pi/2$.}
	\label{fig:fig2} 
\end{figure}

In Fig. \ref{fig:fig2} we show our numerical results for reduced probability distributions for increasing coherent coupling $t_{\rm c}$. 
We observe two main qualitative effects. 
First, the probability distribution spreads in phase space as we increase 
$t_{\rm c}$, in agreement with the expected transition from the PL to the NPL phase. 
Second, we observe a rotation of the probability distribution: at values 
$t_{\rm c} \approx 0$ the two nanolasers are phase-locked to $\theta = 0$, however as we increase $t_{\rm c}$, the distribution rotates to an increasing angle around $\theta = \pi/2$.
This effect can be qualitatively understood with the equations of motion for the coherences of the bosonic modes,
\begin{eqnarray}
\frac{d \langle a_1 \rangle}{d t} &=& 
{\frac{d \langle a_1 \rangle}{d t} } |_{\rm nl} + |\epsilon| e^{i %(\phi_1 - \pi/2)}   - i \ t_{\rm c} \langle a_2 \rangle , \nonumber \\
(\phi_1 - \pi/2)}   + \ t_{\rm c} \langle a_2 \rangle e^{i \pi/2} , \label{eom1} \\
\frac{d \langle a_2 \rangle}{d t} &=& 
{\frac{d \langle a_2 \rangle}{d t} } |_{\rm nl} + |\epsilon| e^{i %(\phi_2 - \pi/2)}   - i \ t_{\rm c} \langle a_1 \rangle .
(\phi_2 - \pi/2)}   + \ t_{\rm c} \langle a_1 \rangle e^{i \pi/2} .
\label{eom2}
\end{eqnarray}
All non-linear and dissipative effects are included in the single nanolaser contribution to the time-evolution of the coherences, $d\langle a_j \rangle/dt |_{\rm nl}$. 
The second and third terms in the r.h.s. of Eqs. (\ref{eom1}, \ref{eom2}) are the external drives and coherent coupling terms, respectively. Note that the external drives -- proportional to $|\epsilon|$ -- contribute with a term that is out of phase by an angle $\pi/2$. The coherent coupling can be understood as an additional external drive with a phase and amplitude determined by the coherence in the nearby cavity mode. 
If we assume that $t_{\rm c}$ is small, then cavities are phase locked to an angle $\theta_j = \phi_j - \pi/2$. However, as we increase $t_{\rm c}$, the photon tunneling process induce an effective external driving, with a phase rotated by an angle $\pi/2$. 
This explains qualitatively the rotation of the probability distribution in Fig. \ref{fig:fig2}.

\begin{figure}[h!]
  \includegraphics[width=\textwidth]{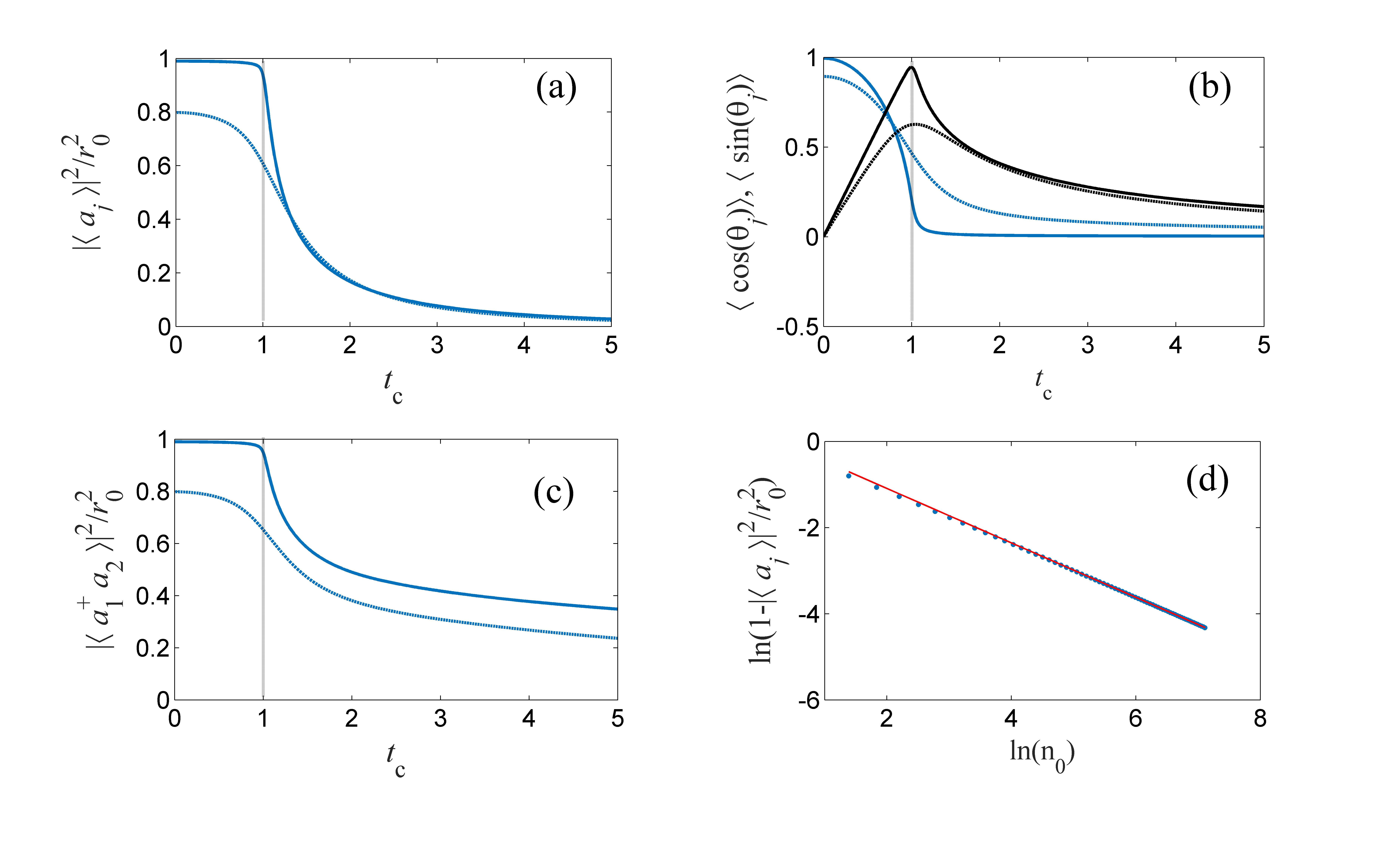}
  \caption{Numerical investigation of the PL-NPL phase transition. All plots are obtained for two nanolasers subjected to an external driving, $\epsilon/r_0  = 1$. 
  In (a-c) we plot the evolution of the coherences with values $D_{\rm p} = 0.2$ (dashed lines) and $D_{\rm p} = 0.005$ (continuous lines). Since $D_{\rm p} \propto 1/n_0$, smaller values of $D_{\rm p}$ are closer to the thermodynamic limit of the problem. (a) Evolution of the coherences in each nanolaser as a function of $t_{\rm c}$. 
  (b) Dependence of the $\cos$ (blue lines) and $\sin$ (black lines) components of the photon coherence as a function of $t_{\rm c}$ 
  (c) Evolution of the two-point photon correlation function as a function of $t_{\rm c}$. 
  (d) Dots are calculated by numerically solving Eq. \ref{PFokker2} at the critical point 
  $t_{\rm c} = |\epsilon|/r_0$, whereas the red line is a linear fit. 
  %We assume values $A/2 = 1$, such that $n_0  = 1/D_{\rm p}$.
  }
\label{fig:fig3}
\end{figure}

To gain a quantitative understanding, we have explored numerically the PL - NPL phase transition as a function of $t_{\rm c}$, see Fig. \ref{fig:fig3}. 
The coherence in each of the nanolasers is calculated with the distribution $P_\theta$ by using the expression,
\begin{equation}
|\langle a_j \rangle|^2 = 
r_0^2 
\left( \langle \cos(\theta_j) \rangle^2  
+ \langle \sin(\theta_j) \rangle^2\right).
\end{equation}

We find in Fig. \ref{fig:fig3} (a) that an abrupt transition between a PL and a UL phase happens at  $t_{\rm c} = 1$, at which the coherences seem to become a non-analytic function of $t_{\rm c}$. The same behaviour is qualitatively observed in the correlation function between cavity modes in Fig. \ref{fig:fig3} (b).
In order to evaluate the finite-size scaling at this critical point we consider a constant coherent coupling $t_{\rm c}$. We then scale $\epsilon$ such that we stay at the critical point $\epsilon/r_0 = \epsilon/\sqrt{n_0} = t_{\rm c}$. 
Increasing the number of photons has thus the effect of decreasing the phase decoherence rate which scales as $D_{\rm p} \propto 1/n_0$. Our numerical results show that the coherence at the critical point follows a power-law dependence,
\begin{equation}
1 - \frac{|\langle a_j \rangle|^2}{n_0} \propto \ n_0^{-\beta},
\end{equation}
and we find the critical exponent $\beta \approx 0.63$ from our numerical calculations (see Fig. \ref{fig:fig3} (d)). 
The critical point is thus a non-analytical point in the thermodynamical limit. 
Finally, by considering the different contributions to the bosonic coherences we have checked the self-rotation of the phase-locking angle induced by the coherent coupling, which is apparent in an increase in the average $\langle \sin(\theta) \rangle$ relative to $\langle \cos(\theta) \rangle$ as we approach the critical point in Fig. \ref{fig:fig3} (b).

\subsection{PL - NPL phase transition with inhomogeneous phases ($\phi_1 \neq \phi_2$)}
\begin{figure}[h!]
  \includegraphics[width = \textwidth]{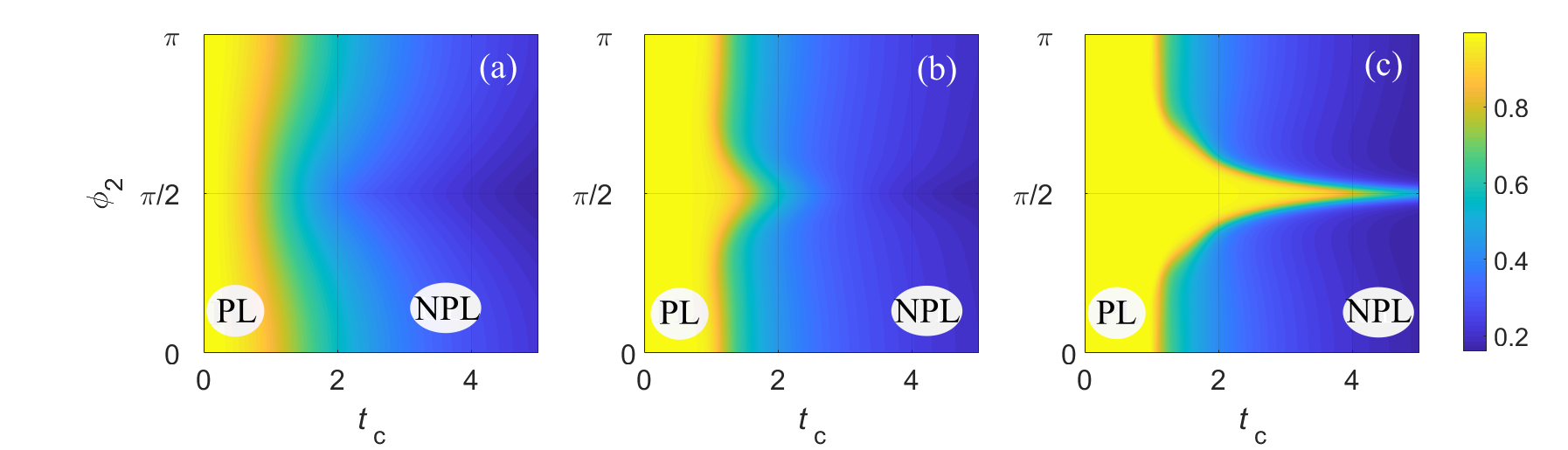}
  \centering
  \caption{We plot the value of the average optical coherence in the nanolaser dimer, $C = \left(|\langle a_1 \rangle|^2/n_0 + |\langle a_2 \rangle|^2/n_0 \right)/2$. 
  This quantity is calculated by using the Fokker-Planck equation, 
  Eq. (\ref{PFokker2}), and the equivalence
  $C = (1/2) \sum_{j=1,2} \left( \langle \cos(\theta_j) \rangle^2 + \langle \sin(\theta_j) \rangle^2 \right)$.
  $\phi_1 = 0$ and $|\epsilon|/r_0 = 1$ in all the plots. 
  (a) $D_{\rm p} = 0.5$, $n_{\rm c} = 20$. 
  (b) $D_{\rm p} = 0.05$, $n_{\rm c} = 100$.
  (c) $D_{\rm p} = 0.005$, $n_{\rm c} = 500$.}
\label{fig:fig4}
\end{figure}

So far we have considered the case of homogeneous driving. An interesting behaviour is found if we study the case of different driving phases $\phi_1 \neq \phi_2$. 
Consider for simplicity that we fix $\phi_1 = 0$ and change $\phi_2$. We plot numerical results in Fig. \ref{fig:fig4}. 
We compare three cases by keeping the same scaling criteria, namely, we fix 
$\epsilon/r_0 = 1$ and calculate the coherence in the nanolaser system as a function of both $t_{\rm c}$ and the angle $\phi_2$.
We observe the PL/UL phase transition at $t_{\rm c} = 1$ for values $\phi_2 = 0, \pi$, in agreement with our results in Fig. \ref{fig:fig3}. However, as we approach the value $\phi_2 = \pi/2$, we observe that the critical $t_{\rm c}$ required to enter into the UL phase increases to values $t_{\rm c} \gg 1$. 
In other words, close to $\phi_2 - \phi_1 = \pi/2$ the PL phase is more robust to a coherent coupling between nanolasers. This effect is more pronounced for large photon numbers and thus lower values of the phase decoherence rate (like in Fig. \ref{fig:fig4} (b) and (c)). We can understand this effect by, again, looking at the Eqs. 
(\ref{eom1},\ref{eom2}). 
If we assume small coherent couplings, then the condition $\phi_1 = 0$ implies, according to Eq. (\ref{eom1}), that the coherent driving of the first cavity mode induced phase locking into $\theta_1 = -\pi/2$. Eq. (\ref{eom2}) then becomes
$\langle \dot{a}_2 \rangle = \langle \dot{a}_2 \rangle_{\rm nl} + 
|\epsilon| e^{i \phi_2 - \pi/2} + t_{\rm c} |\langle a_1 \rangle|$. 
We find that the coherent driving on the second nanolaser and the photon tunneling term have the same phase, only if $\phi_2 = \pi/2$.
This qualitative argument explains the trend observed in the numerical calculation that the PL phase is more resilient to the coherent coupling term, $t_{\rm c}$, when $\Delta \phi = \pi/2$. 
%

%%%%%%%%%%%%%%%%%%%%%%%%%%%%%%%%%%%%%%%%%%%%%%%%%
\section{Photonic Josephson current}
\subsection{Josephson photo-current}
By restricting ourselves to the previously mentioned PL steady-states, interesting phenomena emerge as a result of the interplay between on-site interactions and the coherent tunneling term ($t_{\rm c}$).
In particular, we will show that a photonic analog of the Josephson current is generated between the two nanolasers in the case of a finite difference between the optical phases into which they are locked.

To get a qualitative understanding of this effect, let us examine the evolution of the average photon number, $\langle\dot{n}_{j}\rangle$, in each nanolaser ($j = 1, 2$), which can be calculated by means of the Heisenberg equations obtained with the master equation, Eq. (\ref{adiabatic}).
We distinguish two contributions. 
First, we find a term that is given by the non-linear local dynamics of the nanolasers laser, which, in the limit of large photon numbers, reads
\begin{eqnarray}
\langle\dot{n}_j\rangle |_{\rm nl} 
= 
2 (A - C) \langle n \rangle_j - 2 B_{\rm r} \langle n_j \rangle^2
- 2 \Im \left( \epsilon_j^* \langle a_j \rangle \right) .
\label{lasing.term}
\end{eqnarray}
Note that, apart from corrections arising from the coherent drive, proportional to $\epsilon_j$, the steady-state solution in the lasing phase is 
$\langle n_j \rangle = n_0$, given by Eq. (\ref{n0}). 

Second, we have a contribution arising from the coherent hopping of photons between sites,
\begin{equation}
\langle\dot{n}_1\rangle|_{t_{\rm c}} = 2 t_{\rm c} \Im  \left( \langle a_1^\dagger a_2\rangle \right),
\label{tunneling.term}
\end{equation}
(analogously for $\langle\dot{n}_2\rangle|_{t_{\rm c}}$ ). For weak coherent coupling, $t_{\rm c}$, and small values of $\epsilon_j$, the lowest order approximation of Eqs. (\ref{lasing.term}, \ref{tunneling.term}) can be found by assuming that the effect of the tunneling and coherent drive is negligible in the calculation of the two-point correlation function, 
$\langle a_1^\dagger a_2 \rangle$. In that case, we find an approximate expression,
\begin{equation}\label{Josep}
\langle\dot{n}_1\rangle|_{t_{\rm c}} = - 2 t_{\rm c} n_0 \sin(\phi_1 - \phi_2),
\end{equation} 
which fundamentally indicates a Josephson effect by which a phase difference, $\Delta\phi$, 
of the coherent drivings induces a photo-current between sites. 
This current arises from the coherent nature of the coupling, and it does not appear in the case of a dissipative-mediated coupling \cite{Fernandez18}.

Finally, in the limit of perfect phase-locking, there will be a fixed phase relation between the nanolaser and the coherent drive, such that
\begin{equation}
\Im \langle \epsilon_j^* \langle a_j \rangle \rangle \approx |\epsilon| \sqrt{n_0} .
\end{equation}
We can obtain now an approximate expression for the steady average photon number by adding up (\ref{lasing.term}, \ref{tunneling.term}), and imposing the conditions $\langle\dot{n}_j\rangle |_{\rm nl}+ \langle\dot{n}_j\rangle|_{t}=0$, leading to
\begin{equation}
\langle n_1\rangle_{\rm ss} = n_{\rm 0}  + \frac{ \sqrt{n}_{\rm 0} |\epsilon|}{\kappa(C_{\rm p}-1)} 
-  \frac{t_{\rm c}}{2\kappa^2}\frac{\gamma}{C_{\rm p}} \sin(\phi_1 - \phi_2).
\label{sinusoidal}
\end{equation}
Hence, we find an imbalanced average photon number given by,
\begin{equation} \label{imbalance}
\Delta n_{\rm ss} = 
\langle n_1 \rangle_{\rm ss} - \langle n_2 \rangle_{\rm ss} =  
n_{\rm J}  \sin(\phi_2 - \phi_1), \ \ \
n_{\rm J} = \frac{t_{\rm c}}{\kappa} \frac{\gamma}{C_{\rm p} \kappa} .
\end{equation}
The result in Eq. (\ref{imbalance}) arises from the balance between the individual nanolaser dynamics and the photon tunneling between nanolasers. Note that we have written the solution so as to make explicit the dependence $n_{\rm J} \propto \gamma/\kappa$, which implies that the photon number imbalance can be increased by increasing the incoherent pumping rate. This indicates that the system can act as a good sensor to estimate the phase difference, $\Delta\phi$, 
by simply measuring the imbalance of the average photon number in the steady state of the system. 
%It is now appropriate to discuss the key ingredients used to achieve these results. First, a weak coherent tunneling between two quantum subsystems is required to obtain a Josephson current like (\ref{Josep}). 
%While in a Josephson junction this is typically achieved by means of a thin insulating barrier between two superconducting layers, here we set lattice system. 
%Second, while a superconducting state (or a BEC) assumes the spontaneous symmetry breaking of a gauge symmetry $U(1)$ in the wave function, here we explicitly break the $U(1)$ phase symmetry of the lasing state by applying a couple of coherent periodic drivings which lock the laser phases \cite{Lorenzo17}. Finally, the dissipative non-linear terms given by the local light-matter interaction plus the incoherent dissipation enables us to reach a phase-sensitive steady state with a sensitivity enhanced by an incoherent energy injection.

%%%%%%%%%%%%%%%%%%%%%%%%%%%%%%%%%%%%%%%%%%%%%%%%
\subsection{Numerical investigation of the Josephson current between nanolasers}
So far, our results apply in a limit of strict phase-locking and small coherent couplings. 
However, we have seen in previous sections that the effect of $t_{\rm c}$ is to destroy the coherence in the coupled nanolasers system. 
We have to expect that it is not possible to simply increase the Josephson photocurrent by means of increasing the coupling $t_{\rm c}$, since at some point the system will enter into the NPL, incoherent, steady-state.

To investigate the interplay between Josephson photocurrent and the PL-NPL phase transition we need to carry out numerical calculations that bring us beyond the approximations considered in the previous subsection, in particular beyond the approximation of small $t_{\rm c}$ values. Unfortunately, exact diagonalization of Eq. \ref{Liouvillian} is a numerically challenging task, since we need to deal with a very large Hilbert space. We resort to two approximate methods:

\begin{enumerate}[(a)]
\item {\it Self-consistent Fokker-Planck equation.}
In our first approach we use the inhomogeneous Fokker-Planck equation, Eq. (\ref{PFokker2.0}), which is valid deep in the lasing regime, and we complement it with an approach to account for the evolution of the different photon numbers in each of the two nanolasers. In particular, we rewrite Eqs. (\ref{lasing.term}, \ref{tunneling.term}) in terms of the radial variable of the Fokker-Planck equation,
via the identities 
\begin{eqnarray}
&& \langle a_j \rangle =  r^0_j \langle e^{i \theta_j} \rangle ,
\nonumber \\
&& \langle a^\dagger_j a_{j+1} \rangle =  (r^0_j)^2 \langle e^{i(\theta_{j+1} - \theta_{j})}\rangle .
\end{eqnarray}

The condition $\langle \dot{n}_j \rangle = 0$ reads,
\begin{eqnarray}
(A-C) r_1 - B_{\rm r} (r_1)^3 + |\epsilon| \langle \sin(\phi_1 - \theta_1) \rangle_{P_\theta}
+ t_{\rm c} r_2 \langle \sin(\theta_1 - \theta_2) \rangle_{P_\theta} , 
\nonumber \\
(A-C) r_2 - B_{\rm r} (r_2)^3 + |\epsilon| \langle \sin(\phi_2 - \theta_2) \rangle_{P_\theta}
+ t_{\rm c} r_1 \langle \sin(\theta_2 - \theta_1) \rangle_{P_\theta} ,
\label{self.radial}
\end{eqnarray}
where the angular averages, $\sin(\theta_1 - \theta_2)$ and $\sin(\phi_j - \theta_j)$ have to be evaluated relative to some angular probability distribution, $P_\theta(\theta_1,\theta_2)$. 
We proceed by using the following iterative algorithm:
\begin{enumerate}[(i)]
\item We solve Eq. (\ref{PFokker2}) with initial values $r_1 = r_2 = \sqrt{n_0}$.
\item We use the probability distribution, $P_\theta(\theta_1,\theta_2)$, obtained in step (i) 
to evaluate all the angular terms in Eqs. (\ref{self.radial}), and we solve those equations to obtain new values $r_1$, $r_2$.
\item We solve again the angular equation Eq. (\ref{PFokker2}) with the new values $r_1$, $r_2$ obtained in step (ii)
\item The last two steps are repeated until convergence is reached.
\end{enumerate}

\item {\it Gutzwiller ansatz.} 
To obtain results beyond the semiclassical limit described by the Fokker-Planck equation, we use a Gutzwiller ansatz that is valid in the limit of small photon tunneling rates. The Gutzwiller anstatz approximates the steady state of Liouvillian (\ref{Liouvillian}) by assuming a separable state of the form 
$\rho \approx  \rho_1\bigotimes\rho_2$, where each $\rho_j$ follows a local Liouvillian,
\begin{equation}
\dot{\rho}_j 
= {\cal L}_j \left( \rho_j \right) + i \ t_{\rm c} [a_{j} \alpha^*_{j+1} + a^\dagger_{j} \alpha_{j+1} , \rho_j], 
\end{equation}
where $\alpha_j = {\rm Tr} \left( \rho_j a_j \right)$.  We seek numerically a self-consistent solution of the set of equations. 
Whereas this ansatz neglects quantum and classical correlations between nanolasers, it does allow us to include an exact description of the single nanolaser. 
\end{enumerate}
To summarize the two methods: the self-consistent Fokker-Planck equation method allows us to include correlations between phases, but it relies on the validity of the semiclassical approximation. The Gutzwiller ansatz, on the other hand, allows us to describe exactly the quantum dynamics at the single nanolaser level, but it does not allow to include correlations. In the lasing regime and with small $t_{\rm c}$, the two methods must yield the same results. 

\begin{figure}[h!]
  \includegraphics[width= \textwidth]{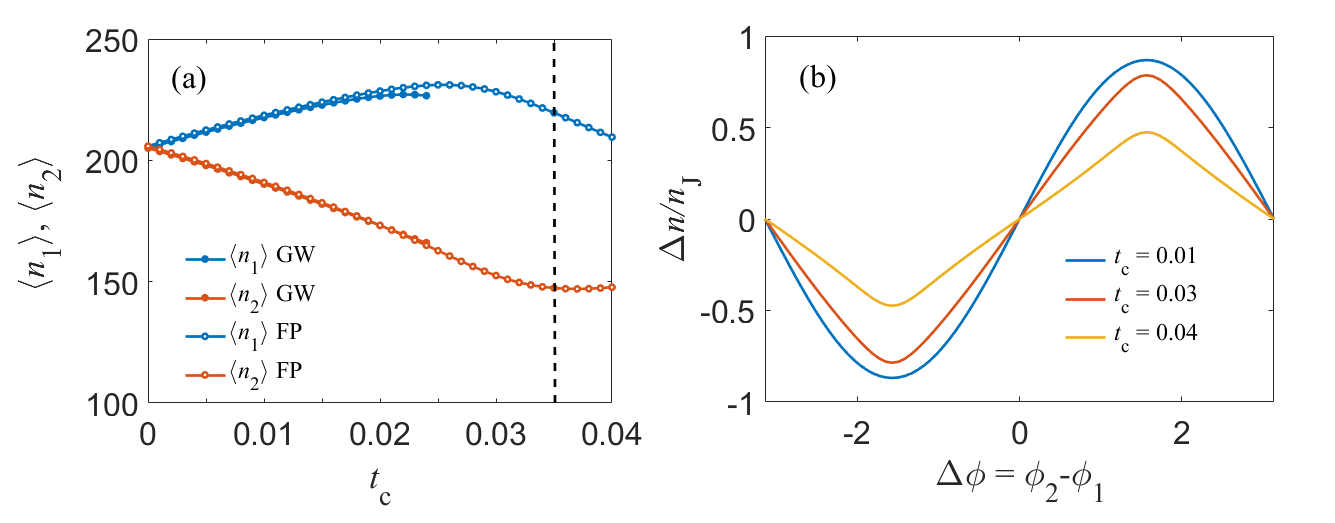}
  \caption{(a) Evolution of the number of photons in each of the nanolasers in the presence of a difference in the phases of the coherent drive, $\phi_1 = 0$, $\phi_2 = \pi/2$. We have considered parameters $\gamma = 3500$, $\kappa = 1$, $\epsilon = 0.3$. The dashed line is the value 
  $t_{\rm c} = \epsilon/\sqrt{n_0}$. This is the point at which, in the thermodynamic limit of an homogeneous system, we find the PL - NPL transition. 
  The filled circles are the calculations with the Gutzwiller ansatz explained in the text. This latter ansatz only works up to values $t_{\rm c} = 0.025$, after which the iterative method to find a self-consistent solution does not converge. Empty points correspond to calculations with the self-consistent Fokker-Planck equation. We see that the photon imbalance saturates at a value of $t_{\rm c}$ consistent with the transition into the NPL steady-state. 
  (b) Relative imbalance between the photon numbers in the two lasers, with the same parameters as in (a). We see that at large values of $t_{\rm c}$ the imbalance departs from the sinusoidal relation, Eq. (\ref{sinusoidal}), predicted in the weak coupling regime.}
  \label{Josephson} 
 \end{figure}

Our numerical calculations are presented in Fig. \ref{Josephson}, where we plot the photon number imbalance  caused by the Josephson photocurrent. 
There is a reasonable agreement between the Gutzwiller ansatz and the self-consistent Fokker-Planck approaches. At large values of the coherent coupling $t_{\rm c}$, the Gutzwiller ansatz does not converge numerically to a steady-state value, and we must assume that neglecting correlations between cavities is not a valid approximation. 
The self-consistent Fokker-Planck equation, on the other hand, is more robust and converges up to higher values of the coupling term, $t_{\rm c}$.
We see in Fig. \ref{Josephson} that the Josephson effect decreases for large values of $t_{\rm c}$, as expected from our discussion on the PL-NPL transition in the previous section.

Figs. \ref{Josephson} (a) and (b) shows that the photocurrent imbalance grows with $t_c$ up to a certain value at which increasing the coupling is detrimental to the coherence between nanolasers, since the system enters into the NPL steady-state.

%%%%%%%%%%%%%%%%%%%%%%%%%%%%%%%%%%%%%%%%%%%%%%%%
\section{The photonic Josephson current as a metrological resource.}
The result in Eq. (\ref{imbalance}) indicates that the nanolaser dimer considered in this work is very sensitive to a phase difference $\phi = \phi_2 - \phi_1$. 
If we calculate the derivative,
\begin{equation}
\frac{\partial \Delta n_{\rm ss} }{\partial \phi}|_{\phi = 0} = n_{\rm J} =
\frac{t_{\rm c}}{\kappa} \frac{\gamma}{C_{\rm p} \kappa},
\label{sensitivity}
\end{equation}
we find that the sensitivity of the system grows linearly with $\gamma/\kappa$, 
which physically can be interpreted in terms of the bosonic amplification of the Josephson photocurrent. 
Furthermore, we also find that the sensitivity is increased as we approach the critical point of the lasing phase transition at $C_{\rm p} = 1$. 

\begin{figure}[h!]
	\centering
  \includegraphics[width=0.6\textwidth]{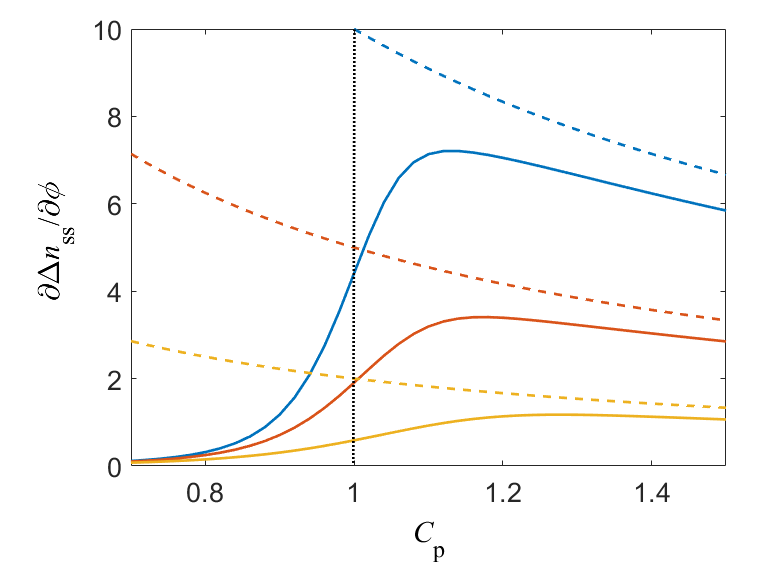}
  \caption{Calculation of the derivative of the photon number imbalance as a function of the cooperativity, $C_{\rm p}$. We use parameters $\gamma =$ $1000$ (blue), $500$ (red), $200$ (yellow), $t_{\rm c} = $0.01, and $\epsilon = $0.3. Continuous lines are the Gutzwiller calculation, whereas the dashed lines are the analytical approximation in Eq. (\ref{sensitivity}).} 
  \label{fig:sensitivity} 
 \end{figure}

We have compared the prediction of the sensitivity around the critical point by using the self-consistent numerical methods introduced in the previous section. We have confirmed that the critical point is indeed the optimal operating point of view for detecting a phase difference, as shown in Fig. \ref{fig:sensitivity}. This result can be qualitatively explained from the dependence of the photon number dynamics on the typical rate scale $A = C_{\rm p} \kappa$, see Eq. \ref{lasing.term}: close to $C_{\rm p}$, the local photon number dynamics slows down, 
thus leading two a stronger effect from the Josephson photocurrent.

Finally, we can also estimate the uncertainty in the value of a phase measurement carried out by the coupled nanolaser system in the quantum limit. The latter assumes that experimental error in photon number measurements is entirely due to the quantum fluctuations of the photon number observable. The latter can be estimated to be,
\begin{equation}
    \Delta (n_1 - n_2)_{\rm ss}  \approx \   2  \ \left( \frac{1}{2} \frac{\gamma}{\kappa} \right)^{1/2},
\label{err.difference}
\end{equation}
where we have used the calculation of photon number fluctuations obtained with the single nanolaser radial Fokker-Planck equation, Eq. (\ref{Dn.Gaussian}). 
Eq. (\ref{err.difference}), together with Eq. (\ref{sensitivity}) allows us to estimate the uncertainty in the estimation of a small phase difference, $\delta \phi$, from a measurement of the difference in photon numbers in the coupled nanolasers,
\begin{equation}
    \delta \phi = 
    \frac{\Delta (n_1 -n_2 )_{\rm ss}}
    {{\frac{\partial \Delta n_{\rm ss} }{\partial \phi}}|_{\phi = 0} } = 
    \sqrt{2} \ C_{\rm p} \frac{\kappa}{t_{\rm c}} 
    \left( \frac{\gamma}{\kappa} \right)^{-1/2}
\end{equation}
The last expression shows that the error grows with $C_{\rm p}$, and it scales like $\delta \phi \propto 1/\sqrt{\gamma/\kappa} \approx 1/\sqrt{n}$. This expression shows that the amplification effect survives in the quantum limit, in which the accuracy can be enhanced by increasing the number of photons, for example by increasing the excitation rate of the qubits, $\gamma$.

%%%%%%%%%%%%%%%%%%%%%%%%%%%%%%%%%%%%%%%%%%%%%%%%%%%%
\section{Conclusions \& Outlook}
We have presented a theoretical study of two nanolasers coupled by a photon tunneling term. We have arrived to two main conclusions. The first is that a photon tunneling term is a source of decoherence which can ultimately destroy the phase locking of each individual nanolaser to an external driving field. The second conclusion is that, in a limit of small photon tunneling rates, a photonic Josephson effect is induced that can be used to measure the phase difference between external fields.

The model of interferometer proposed in this article may be implemented in circuit QED systems. In this experimental platform, single-qubit nanolasers have already been demonstrated \cite{astafiev07nat}, and schemes for controlling the properties of single-qubit laser light, including the generation of entangled states of light, have been proposed \cite{navarrete14prl}. Photon tunneling is actually the main mechanism that couples microwave cavities in circuit QED \cite{houck12natphys}, thus making this system an ideal experimental platform for implementing lattices of coupled nanolasers.   
Our ideas can also be translated to the realm of phononics, for example in trapped ion setups. Here, the coupling between internal electronic states acting as quibts and the vibrational degrees of freedom can be controlled with optical lasers. 
Single ion phonon lasing has actually been experimentally demonstrated \cite{vahala09}. 
Vibrational modes of coupled trapped ions can often be described in terms of local phononic modes, with the Coulomb interaction between ions inducing a phonon hopping process \cite{Porras04prl2}, an effect that has been experimentally observed \cite{Urabe2012pra, Monroe18prl}. Trapped ions actually offer the possibility to study few-mode coupled lasers or include highly controllable qubit-phonon interactions  \cite{Porras12prl} .

Our work could be extended to larger lattice sizes, something that could potentially allow high sensitive estimation of phase gradients. Additionally, the simultaneous implementation of coherent couplings and dissipative-mediated couplings \cite{Fernandez18} is expected to exhibit interesting dissipative phase transitions of photonic phases, also interesting for studying quantum synchronization.
An intriguing possiblity here is the investigation of non-reciprocal couplings and topological effects, as well as topological amplification \cite{Porras19prl}.
Finally, these results invite to study further Josepshon effects or configurations, like the a.c. Josepshon effect or the SQUID \cite{jaklevic164}.
 
\section{Acknowledgments.-} 
Funded by the People Programme (Marie Curie Actions) of the EU’s Seventh Framework Programme under REA Grant Agreement no: PCIG14-GA-2013-630955, and project PGC2018-094792-B-I00 (MCIU/AEI/FEDER, UE).

\appendix

%%%%%%%%%%%%%%%%%%%%%%%%%%%%%%%%%%%%%%%%%%%%%%%%%%%%%%%%%%%%%%
\section{Adiabatic elimination}\label{App:AppendixB}
In this appendix we show how the adiabatic elimination of the qubits leads to 
Eq. (\ref{adiabatic}).
For this we employ a generalization of the procedure discussed in Refs. \cite{Lorenzo17, Fernandez18}. 
Since we assume that photon tunneling rates smaller than local energy scales, we consider the single nanolaser case for this derivation. 

Firstly, we trace over the qubits from the master equation. Since the effect of the photon decay, $\kappa$, is not affected by the adiabatic elimination procedure, we set $\kappa = 0$, and reintroduce it later in the calculation,
\begin{eqnarray}
\dot{\rho}_{\rm f} = 
\langle e | \dot{\rho} | e \rangle =
-i g (a \rho^{ge} +a^{\dag} \rho^{eg} -\rho^{ge} a -\rho^{eg} a^{\dag}),
\label{rhof}
\end{eqnarray}
where $\rho_{\rm f}$ is the reduced density matrix of the field after tracing over the qubit states, 
$\rho_{\rm f} = {\rm Tr}_q \{\rho\} = \langle g | \rho | g \rangle + \langle e | \rho | e \rangle$, and we employ the notation $\rho^{ge} = \langle g | \rho | e \rangle$.
To obtain a closed equation for $\rho_{\rm f}$ we need to eliminate the operators $\rho^{ge}$, $\rho^{eg}$ from Eq. (\ref{rhof}). 
The corresponding equations of motion for these operators are,
\begin{equation}
\dot{\rho}^{ge} = - i g(a^{\dag}\rho^{ee} -\rho^{gg} a^{\dag}) - \gamma \rho^{ge}  ,
\label{rhoge}
\end{equation}
The operators $\rho_{ge}$ and $\rho_{eg}$ may now be adiabatically eliminated (in the limit  $\gamma \gg \kappa, g, |\epsilon|$) from Eq. (\ref{rhof}) by taking $\dot{\rho}_{ge}\approx 0$ in Eq. (\ref{rhoge}) and substituting their steady-state solutions,
\begin{equation}
\rho^{ge} = -i \frac{g}{\gamma}(a^{\dag} \rho^{ee} - \rho^{gg} a^{\dag}) . 
\label{coherences}
\end{equation}
Now we can use Eq. (\ref{coherences}) and insert it into Eq. (\ref{rhof}) and get,
\begin{eqnarray}
\dot{\rho}_{\rm f} = 
- \frac{g^2}{\gamma} a a^\dagger \rho^{ee}   
+ \frac{g^2}{\gamma} a \rho^{gg} a^\dagger +
\frac{g^2}{\gamma} a^\dagger \rho^{ee} a -
\frac{g^2}{\gamma} a^\dagger a \rho^{gg}  + {\rm H.c.} 
\label{rhof2}
\end{eqnarray}

We need now equations of motion for $\rho_{gg}$ and $\rho_{ee}$, which can be derived again from Eq. (\ref{Liouvillian}), leading to
\begin{eqnarray}
\dot{\rho}^{ee} &=&-i g(a \rho^{ge} - \rho^{eg} a^{\dag}) +2\gamma \rho^{gg} \label{rhoeedot}\\ 
\dot{\rho}^{gg} &=&-i g(a^{\dag} \rho^{eg} - \rho^{ge} a) -2\gamma \rho^{gg} \label{rhoggdot}
\end{eqnarray}
We may now reach a perturbative solution to the steady-states of 
Eqs. (\ref{rhoeedot}, \ref{rhoggdot}) in terms of the field density matrix $\rho_{\rm f}$ by adiabatically eliminating $\rho^{gg}$ and $\dot{\rho}^{gg}$. This is done by taking $\dot{\rho}^{gg} \approx 0$ in Eq. (\ref{rhoggdot}), 
\begin{equation}
\rho^{gg} = -\frac{ig}{2\gamma}(a^{\dag} \rho^{eg}-\rho^{ge} a)
=\frac{g^2}{2\gamma^2}(2a^{\dag} \rho^{ee} a-a^{\dag} a \rho^{gg} - \rho^{gg} a^{\dag} a) .
\label{rhogg}
\end{equation}
The ground state population of each qubit is expected to be negligible as a result of the fast pumping of the qubits. Thus, in first order we can assume $\rho^{gg} \approx 0$ and 
$\rho^{ee} =\rho_{\rm f}-\rho^{gg} \approx \rho_{\rm f}$. A second order approximation is achieved by inserting this first order approximation into Eq. (\ref{rhogg}), hence
\begin{eqnarray}
\rho^{gg} &=& \frac{g^2}{\gamma^2} a^{\dag} \rho_{\rm f} a \label{rhogg2} \\ 
\rho^{ee} &=& \rho_{\rm f} - \rho^{gg} = \rho_{\rm f} - \frac{g^2}{\gamma^2} a^{\dag} \rho_{\rm f} a 
\label{rhoee}.
\end{eqnarray}
The desired closed equation for $\rho_{\rm f}$ is accomplished by plugging Eqs. (\ref{rhogg2}, \ref{rhoee}) into Eq. (\ref{rhof2}), leading to
\begin{eqnarray}
\dot{\rho}_{\rm f} &=& 
\frac{g^2}{\gamma}( 2 a^{\dag} \rho_{\rm f} a - aa^{\dag} \rho_{\rm f}- \rho_{\rm f} aa^{\dag}) .
\nonumber \\
&+& 
\frac{2g^4}{\gamma^3}(  aa^{\dag} \rho_{\rm f} a a^{\dag}-{a^{\dag}}^2 \rho_{\rm f} a^2) + 
\frac{2g^4}{\gamma^3}(  a^{\dag} \rho_{\rm f} a )
\label{adiabaticEquation}
\end{eqnarray}

We can check, by using the commutation relations of $a$ that the equation can be written in Lindbladt form,
\begin{eqnarray}
\dot{\rho}_{\rm f} &=& 
A \left( 2 a^{\dag} \rho_{\rm f} a -aa^{\dag} \rho_{\rm f}- \rho_{\rm f} aa^{\dag} \right) \nonumber \\ 
&+& 
B \left(
 aa^{\dag} \rho_{\rm f} a a^{\dag} - \frac{1}{2} a a^{\dag} a a^{\dag} \rho_{\rm f} - \frac{1}{2} \rho_{\rm f} a a^{\dag} a a^{\dag} 
 \right)
\nonumber \\
&-& 
B \left(  
{a^{\dag}}^2 \rho_{\rm f} a^2 - \frac{1}{2} a^2 {a^{\dag}}^2 \rho_{\rm f} 
- \frac{1}{2} \rho_{\rm f} a^2 {a^{\dag}}^2  
\right) 
 \nonumber \\
&+& 
B \left( a^{\dag} \rho_{\rm f} a  - \frac{1}{2} a a^{\dag} \rho_{\rm f} a - \frac{1}{2} \rho_{\rm f} a a^{\dag}  \right).
\label{adiabaticEquation2}
\end{eqnarray}
with definitions $A g^2/\gamma$, 
$B 2 g^4/\gamma^3$. If we include now the photon decay term, we finally get to Eq. (\ref{adiabatic}).
%

%
%

%
%%%%%%%%%%%%%%%%%%%%%%%%%%%%%%%%%%%%%%%%%%%%%%%%%%%%%%%%%%%%%%%%%%%%%
%%%%%%%%%%%%%%%%%%%%%%%%%%%%%%%%%%%%%%%%%%%%%%%%%%%%%%%%%%%%%%%%%%%%%%%
%%%%%%%%%%%%%%%%%%%%%%%%%%%%%%%%%%%%%%%%%%%%%%%%%%%%%%
\section{Renormalization of the adiabatic equation}\label{App:AppendixC}
Eq. (\ref{adiabaticEquation}) is a perturbative description in the limit of adiabatic elimination of the fast qubit dynamics. 
This equation is strictly valid below the critical point, $C_{\rm p}<1$, and slightly above it, $C_{\rm p}\gtrsim 1$ \cite{Lorenzo17}, which implies a severe limitation in the range of applicability of our results.

For the single qubit laser the average number of bosons in the steady state predicted by 
Eq. (\ref{adiabatic}) is,
\begin{equation}
n_{\rm ph} \approx \frac{A - B}{C} = \frac{1}{2} \frac{C_{\rm p} - 1}{C_{\rm p}^2} \frac{\gamma}{\kappa},
\label{nph}
\end{equation}
which tends to zero for large values of the pumping parameter. 
Eq. (\ref{nph}) can be obtained by finding the steady-state value of the photon number operator and neglecting photon number fluctuations. This prediction differs from the mean field result \cite{Lorenzo17},
\begin{equation}
n_{\rm mf} = \frac{1}{2} \frac{C_{\rm p} - 1}{C_{\rm p}} \frac{\gamma}{\kappa}. \label{nMF}
\end{equation}
The two expressions agree only close to $C_{\rm p} = 1$. Still, one may think of a renormalization procedure such that allows us to perform a summation of the neglected terms in the perturbative series with respect to $(g/\gamma)^2$. 
The root of the perturbative nature of Eq. (\ref{adiabaticEquation}) is given by the truncation performed in Eq. (\ref{rhogg2}). 
We search for a new adequate parameter $\beta$ that takes into account the remaining terms of the series such that,
\begin{equation}
\rho^{gg} = 
\alpha \frac{g^2}{\gamma^2} a^\dagger \rho_{\rm f} a .
\label{rhoggRenorm} 
\end{equation}

We discuss now a procedure to compute $\beta$ by using an exact relation held in the steady state which is easily inferred from the Eq. (\ref{Liouvillian}) of the Letter. 
Let us calculate the Heisenberg equation of motion for the observable
$N = a^{\dag} a + \sigma^z /2$, which reads, for the single site case,
\begin{figure}[h!]
	\includegraphics[width=0.95\textwidth]{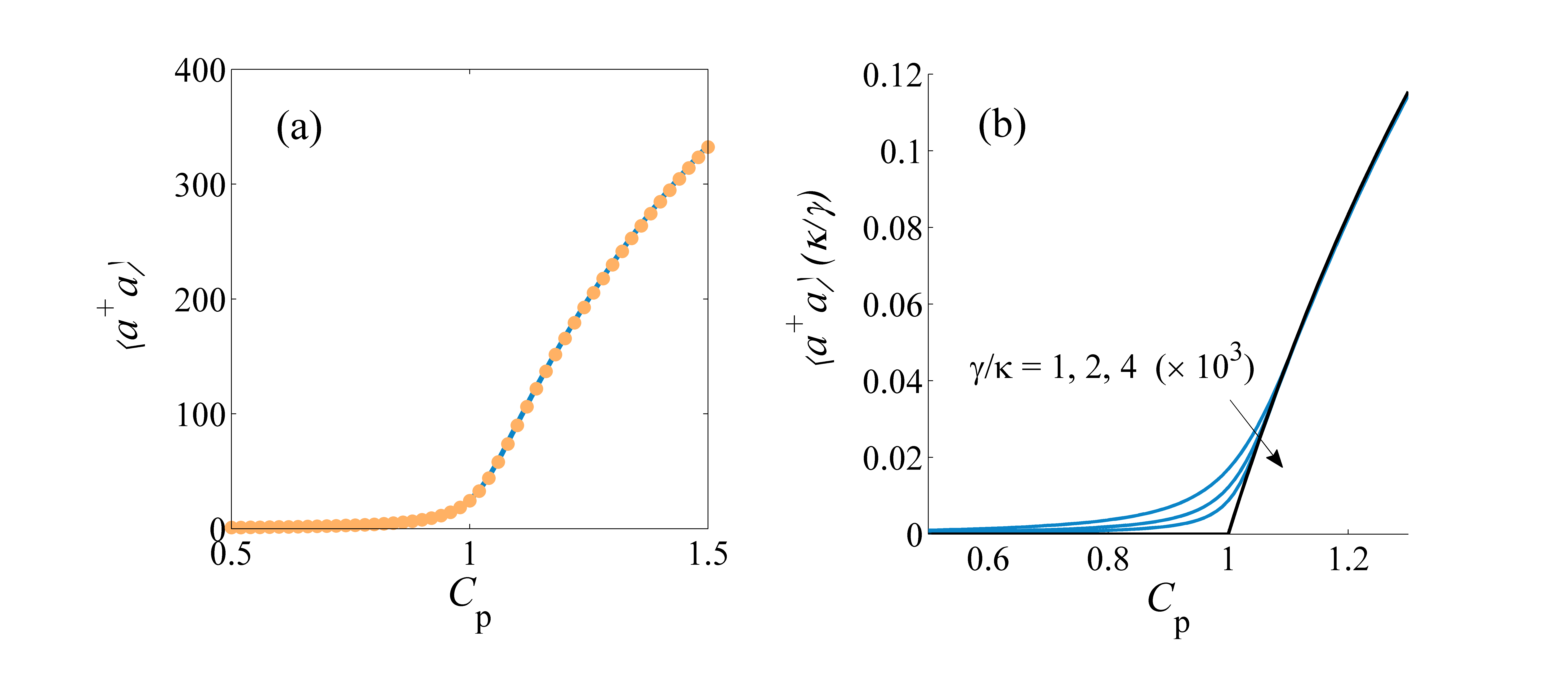}
	\caption{(a) Comparison between the number of photons in the steady state of a single nanolaser predicted by the full qubit-photon Liovillian in Eq. (\ref{Liouvillian}) (blue line) with the result obtained after tracing out the qubit (single nanolaser version of Eq. (\ref{adiabatic})) with the substituion $B \to B_{\rm r}$ (orange points). We have considered 
	$\gamma = 2 \times 10^3$, $\kappa = 1$. 
	(b) Photon number curves calculated with the single nanolaser version of Eq. (\ref{adiabatic}) and the substitution $B \to B_{\rm r}$, showing the scaling with $\gamma/\kappa$ and the convergence of the curves towards the mean-field solution.}
	\label{comparison} 
\end{figure}

\begin{equation}
\frac{d\langle  N \rangle}{dt}
=
- 2 \kappa \langle a^{\dag} a \rangle 
- \gamma \left(\langle\sigma^z \rangle - 1 \right). \label{NumberExcit}
\end{equation}
In the steady state, $\frac{d\langle N\rangle}{dt}=0$, which implies
\begin{equation}
1 -\langle \sigma^z \rangle = \frac{2\kappa}{\gamma} \langle n \rangle ,
\label{SteadyCondition}
\end{equation}
where $n = a^{\dag} a $. Notice that this is an exact relation. 
On the other hand, by taking traces in both sides of (\ref{rhoggRenorm}), we find the following equation
\begin{equation}
2 {\rm Tr}(\rho^{gg})
=
1-\langle \sigma^z \rangle
= 
2 \beta \frac{g^2}{\gamma^2}(\langle n \rangle + 1). \label{RenormRelation}
\end{equation}
We equate now Eqs (\ref{SteadyCondition}) and (\ref{RenormRelation}), and take the limit 
$\langle n \rangle \gg 1$, and get,
\begin{equation}
\beta = \frac{1}{C_{\rm p}}.
\end{equation}
This eventually leads to a renormalization of the parameter $B$ that amounts to $B\rightarrow 
B/C_{\rm p}\equiv B_{\rm r}$. The latter leads to a correction of Eq. (\ref{nph}), such that
\begin{equation}
    n_{\rm ph} \to \frac{A - B_{\rm r}}{C} = n_{\rm mf},
\end{equation}
so that the number of photons finally actually agrees with the mean result. 
Furthermore, in Fig. \ref{comparison} (a) we present numerical results that show that the photon number calculated with Eq. (\ref{adiabatic}) agrees very well with an exact single qubit laser calculation using Eq. (\ref{Liouvillian}). 
In Fig. \ref{comparison} (b) we show calculations carried out with the renormalized $B_{\rm r}$ parameter, that show that the number of photons curves converge to the mean-field result for large values of $\gamma/\kappa$.

Note that for smaller values of $\langle n \rangle$ it is possible to make an $n$-dependent definition of $\beta$. One could then carry out an iteration procedure 
(e.g. estimate $\langle n \rangle$ with $\beta = 1/C_p$, then use this value to re-calculate $\beta$, calculate $\langle n \rangle$ again, and so on until reaching convergence). 
The added benefit of this is, however, not very high, since our renormalization procedure works very well in practice.

%%%%%%%%%%%%%%%%%%%%%%%%%%%%%%%%%%%%%%%%%%%%%%%%%%%%%%%%%%%%%%%%%%%%%
%%%%%%%%%%%%%%%%%%%%%%%%%%%%%%%%%%%%%%%%%%%%%%%%%%%%%%%%%%%%%%%%%%%%%%%
%%%%%%%%%%%%%%%%%%%%%%%%%%%%%%%%%%%%%%%%%%%%%%%%%%%%%%
\section{Fokker-Planck equation}\label{App:AppendixD}

\subsection{Derivation of the Fokker-Planck equation}

Let us sum up the derivation of the Fokker-Planck equations used in this work. 
Recall that we employ the \textit{Glauber-Sudarshan P} representation of the density matrix \cite{Mandel95}, 
defined by
\begin{equation}\label{coherentP}
\rho(t)=\int d^2\alpha P(\alpha,\alpha^*,t) |\alpha\rangle\langle \alpha |, 
\end{equation}
where $|\alpha\rangle$ is the coherent state $|\alpha\rangle=\exp{(\alpha a^{\dag}-\alpha^* a)}|0\rangle$. We transform the master equation (\ref{adiabaticEquation}) with the help of the following relations for coherent states,
\begin{eqnarray}
	a|\alpha\rangle\langle \alpha |&=&\alpha |\alpha\rangle\langle \alpha |, 
	\nonumber \\ 
	|\alpha\rangle\langle \alpha | \adag &=&\alpha^* |\alpha\rangle\langle \alpha |, 
	\nonumber \\
	\adag |\alpha\rangle\langle \alpha | &=&\left( \frac{\partial}{\partial\alpha} +\alpha^* \right)  |\alpha\rangle\langle \alpha | , 
	\nonumber \\ 
	|\alpha\rangle\langle \alpha |a &=&\left( \frac{\partial}{\partial\alpha^*} +\alpha \right)  |\alpha\rangle\langle \alpha | . \label{alphaderiv}
\end{eqnarray}
With Eqs. (\ref{alphaderiv}) we can write the master equation as an equation of motion for $P(\alpha,\alpha^*,t)$, after an integration by parts with the assumption of zero boundary conditions at infinity. 
This procedure is simplified in the limit 
$|\alpha|^2\gg 1$, and $g \ll \gamma$
in which we drop any contribution smaller than $B |\alpha|^2 \alpha$, as $B$ happens to be a very small coefficient compared to $A$, $B/A\propto(g/\gamma)^2 \ll 1$. 
In doing so, we arrive at the Fokker-Planck equation claimed in the Letter,
\begin{eqnarray} \label{PFokker1A}
\frac{\partial P}{\partial t}&=& + 2A\sum_{j=1,2}\frac{\partial^2P}{\partial \alpha_j\partial\alpha_j^*} \\
 &-&\sum_{j=1,2}\frac{\partial}{\partial \alpha_j} [(A-C-B_{\rm r}|\alpha_j|^2)\alpha_j  
 -i t_{\rm c} \alpha_{j+1}-i\epsilon_j] P + {\rm C.c.}  \nonumber
\end{eqnarray}

Our equation of motion can be written in polar coordinates with the aid of the equivalences,
\begin{eqnarray}
\alpha_j &=& r_j e^{i\theta_j}  , \ \ \alpha_j^* = r_j e^{-i\theta}, 
\nonumber \\
\frac{\partial}{\partial\alpha_j}
&=&\frac{1}{2}e^{(-i\theta_j)}
\left(\frac{\partial}{\partial r_j}
-\frac{i}{r_j}
\frac{\partial}{\partial \theta_j}\right), 
\nonumber\\
\frac{\partial}{\partial\alpha_j^*}
&=&\frac{1}{2}e^{(i\theta_j)}
\left(\frac{\partial}{\partial r_j}+\frac{i}{r_j}\frac{\partial}{\partial \theta_j}\right).
\end{eqnarray}
Eq. (\ref{PFokker1A}) then reads,
\begin{eqnarray}
\label{FokkerPolarA}
& & \frac{\partial P}{\partial t}= \frac{\prod_j\partial R_j}{\partial t}P'+\frac{\partial P'}{\partial t}\prod_j R_j= +\frac{A}{2}\sum_j\frac{\partial^2}{\partial \theta_j^2}P- \nonumber\\
&-& \sum_j\left\{\frac{1}{r_j}\frac{\partial}{\partial r_j}\left[r_j^2(A-C-B_{\rm r}r_j^2)P\right] +\frac{A}{2}\left[\frac{\partial^2}{\partial r_j^2}+\frac{1}{r_j^2}\frac{\partial}{\partial r_j}\right]P\right\}+\nonumber\\
&+& |\epsilon|\sum_j\sin(\theta_j-\phi)\frac{\partial P}{\partial r_j}+\sum_j\frac{|\epsilon|}{r_j}\cos(\theta_j-\phi)\frac{\partial P}{\partial\theta_j}+\nonumber\\
&+& t\sum_{j}r_{j+1}\sin(\theta_{j+1}-\theta_j)\frac{\partial P}{\partial r_j}-t\sum_{j}\frac{r_{j+1}}{r_j}\cos(\theta_j-\theta_{j+1})\frac{\partial P}{\partial\theta_j}.
\end{eqnarray}
This is a very complicated equation, whose analytical or even numerical solution if very challenging. We are going to see below how a simplification can be justified in the lasing regime.

\subsection{Single nanolaser Fokker-Planck equation}
Before addressing the case of two coupled nanolasers, let us consider the solution of a single nanolaser Fokker-Planck equation without any external drive ($\epsilon = 0$). 
In this case, 
$P(r,\theta) = R(r) P_\theta(\theta)$, 
with $P_\theta = 1/(2 \pi)$.
The radial function, $R(r)$, satisfies the differential equation,
\begin{equation}
\frac{d R}{d t}
= \frac{1}{r}\frac{\partial}{\partial r}
\left( r^2(A-C-B_{\rm r} r^2) R \right) 
+ \frac{A}{2}\left(\frac{\partial^2}{\partial r^2}+\frac{1}{r}\frac{\partial}{\partial r} \right) . 
    \label{radial.FP}
\end{equation}
The steady-state solution of Eq. (\ref{radial.FP}) with $dR/dt = 0$ is,
\begin{equation}
R(r) = \frac{1}{N} e^{-\frac{B_\rr}{2 A} r^4 + \frac{A - C}{B} r^2} =
\frac{1}{N} e^{ \frac{\kappa}{\gamma} r^4 + \frac{C_{\rm p} -1 }{C_{\rm p}} r^2},
\end{equation}
where $N = \int_0^\infty dr r R(r)$, is a normalization constant.

In the lasing regime ($C_{\rm p} > 1$) and in the limit of large photon numbers ($\gamma/\kappa > 1$), the steady-state radial probability function  takes the form,

\begin{equation}
    R(r)  = \frac{1}{\sqrt{2\pi}r_0 \sigma} e^{\frac{-(r-r_0)^2}{2 \sigma^2}} ,
\end{equation}
with the constants
\begin{eqnarray}
(r_0)^2 &=& \frac{A-C}{B_{\rm r}} = \frac{1}{2} \frac{C_{\rm p}-1}{C_{\rm p}} \frac{\gamma}{\kappa} , \nonumber \\
\sigma^2  &=& \frac{1}{4} \frac{A}{A - C} = \frac{1}{4} \frac{C_{\rm p}}{C_{\rm p} - 1}.
\end{eqnarray}
The radial distribution can be used to calculate the mean and variance of the photon number,
\begin{eqnarray}
&& \langle n \rangle = (r_0)^2,   \label{n.Gaussian} \\ 
&& \Delta^2 n = \langle n^2 \rangle - \langle n \rangle^2 
\approx \frac{1}{2} \frac{\gamma}{\kappa} \label{Dn.Gaussian} .
\end{eqnarray} 
Note that Eq. (\ref{n.Gaussian}) agrees with the mean-field calculation ($n_{\rm mf} = (r_0)^2$). Also, Eq. (\ref{Dn.Gaussian}) agrees with the calculation of the variance from a coherent state in the limit of $C_{\rm p} \gg 1$, for which we recover the well-known relation $\Delta^2 n = \langle n\rangle$ \cite{WallsMilburn}. 

\subsection{Reduced Fokker-Planck equation for two nanolasers}

In the lasing regime, we can simplify  Eq. (\ref{PFokker1A}) by assuming that the radial variables are settled around their steady-state mean values $r_j\approx r^0_j$, and 
$P(\alpha,\alpha')$ can be well approximated by a factorized form, 
\begin{equation}
P(\mathbf{r},\mathbf{\theta}) = R(r_1)R(r_2)P_\theta (\theta_1,\theta_2), 
\label{factorize.P}
\end{equation}
where each $R(r_j)$ is a properly normalized Gaussian function around $r^0_{j}$, 
\begin{equation}\label{radial}
R(r_j)=\frac{1}{N_j}\exp{\left(-\frac{(r_j-r^0_{j})^2}{2\sigma^2}\right)} ,
\ \  N_j = \sqrt{2 \pi} r^0_j \sigma
\end{equation}
Eq. (\ref{factorize.P}) is valid as long as the phase variable evolves on much slower time scales as the radial variables. 
The typical times scales for phase $\tau_\phi$ and radial $\tau_{r}$ variables can be estimated, in the lasing regime and assuming values of $r_j$ close to $r^0_j$, from Eq. (\ref{FokkerPolarA}),
\begin{eqnarray}
1/\tau_\theta &=& \frac{A}{2 \langle n \rangle} = \kappa C_{\rm p}, \nonumber \\
1/\tau_r &=& \frac{A}{2} .
\end{eqnarray}
Additionally, the relaxation time associated to the photon number variable can also be estimated by means of the Heisenberg equations, see Eq. (\ref{lasing.term}).  We find, thus, that $\tau_{\rm r} \approx n \tau_{\rm \phi}$, such that in the large photon number limit, the nanolaser radial coordinate can be considered to relax very fast compared with the phase variable. This behaviour is consistent with other theoretical descriptions of the laser's dynamics \cite{WallsMilburn}.

If we now integrate both sides of Eq. (\ref{FokkerPolarA})  in the radial variables $\int_0^{\infty}\vec{r}d\vec{r}$, the result is an equation of motion for the angular part. 
In this derivation it is important to notice that the terms 
$\partial R_j / \partial t $ can be cancelled together with other terms by using the identity Eq. (\ref{radial.FP}). Also, the integration of the first derivative $\partial_{r}R$ is eliminated through the relation,
\begin{eqnarray}
\int_0^{\infty}r_jdr_j\partial_rR&=&-\frac{1}{N_j}\int_0^{\infty}r_jdr_j\frac{(r_j-r^0_j)}{\sigma^2}\exp\left(-\frac{(r_j-r^0_j)^2}{2\sigma^2}\right)=\nonumber\\
&=&-\frac{1}{N}\int_{-r_0}^{\infty}(r'+r_0)dr'\frac{(r')}{\sigma^2}\exp\left(-\frac{(r')^2}{2\sigma^2}\right)\approx\nonumber\\
&\approx& -\frac{1}{N_j}\int_{-\infty}^{\infty}d{r'}_j\frac{({r'}_j^2)}{\sigma^2}\exp\left(-\frac{(r')^2}{2\sigma^2}\right)=\nonumber\\
&=&
-\frac{1}{N}\sqrt{2\pi}\sigma=-\frac{1}{r^0_j}.
\end{eqnarray}

After grouping terms, the resulting equation adopts the form claimed in equation (6) of the Letter,
\begin{eqnarray} \label{PFokker2A}
\frac{\partial P'}{\partial t}&=& +\frac{A}{2}\sum_{j=1,2}\frac{1}{n^0_j}\frac{\partial^2P'}{\partial \theta_j^2}  \\
+\sum_{j=1,2} & &\frac{\partial}{\partial \theta_j}\left((t\frac{r^0_{j+1}}{r^0_j}\cos(\theta_{j+1}-\theta_j)+\frac{|\epsilon|}{r^0_j}\cos(\theta_j-\phi_j))P'\right) \nonumber,
\end{eqnarray}
Notice that this a sort of quantum version of the stochastic Kuramoto model, that is, it represents two Fokker-Planck equation coupled by a term that was originated by coherent photon tunneling. 

\section*{References}
\bibliographystyle{iopart-num}
\bibliography{biblio}
\end{document}